\documentclass{sigplanconf}
\usepackage[utf8]{inputenc}

\usepackage{amsmath,amsfonts,amstext,amssymb,txfonts,pifont,url}

\renewcommand{\paragraph}[1]{\vspace{\parsep} \noindent {\bf #1}~}
\def\maxlive{\mbox{Maxlive}}
\newcommand{\Cqfd}{\hfill $\square$
 \smallskip
}
\newcommand{\cplx}{\mbox{\it O}}

\ifx\usetikz\undefined\else
\usepackage{tikz}
\usetikzlibrary{arrows,shapes}

\pgfdeclarelayer{background}
\pgfsetlayers{background,main}

\tikzstyle{vertex}=[circle,fill=black!50,minimum size=20pt,inner sep=0pt]
\tikzstyle{vsimpl}=[circle,fill=black!25,minimum size=20pt,inner sep=0pt]
\tikzstyle{vertb}=[circle,fill=blue!25,minimum size=20pt,inner sep=0pt]
\tikzstyle{vertr}=[circle,fill=red!25,minimum size=20pt,inner sep=0pt]
\tikzstyle{vertg}=[circle,fill=green!25,minimum size=20pt,inner sep=0pt]
\tikzstyle{interf} = [draw,thick,-]
\tikzstyle{affinity} = [draw,thick,dotted,-]
\tikzstyle{selected edge} = [draw,line width=5pt,-,red!50]
\tikzstyle{ignored edge} = [draw,line width=5pt,-,black!20]
\fi

\usepackage{texgraphicx}
\usepackage{epsfig}
\usepackage{color}
\graphicspath{{fig/}}

\usepackage{endnotes}

\newtheorem{theorem}{Theorem}

\newcommand{\PROBLEM}[3]{
\medskip
\centerline{\fbox{
\begin{minipage}{0.45\textwidth}
\paragraph{Problem: \textsc{#1}}\\
{\bf Instance} #2\\
{\bf Question} #3
\end{minipage}
}}\medskip}

\newcommand{\PROBLEME}[4]{
\medskip
\centerline{\fbox{\begin{minipage}{0.45\textwidth}
\paragraph{Problem: \textsc{#1}}\\
{\bf Instance} #2\\
{\bf Question} #3\\
{\bf Other instances} #4
\end{minipage}}}\medskip}

 \newenvironment{proof}[1]%
 {\paragraph{Proof:}{#1}}%
 {\Cqfd}

\def\P{\mbox{$\mathbb{P}$}}
\def\NP{\mbox{$\mathbb{NP}$}}

\begin{document}

\conferenceinfo{LCTES'07} {June 13--16, 2007, San Diego, California, USA.}
\copyrightyear{2007}
\copyrightdata{978-1-59593-632-5/07/0006} 

\titlebanner{DRAFT --- Do not distribute}        
\preprintfooter{Our spill paper for LCTES'07}    

\title{On the Complexity of Spill Everywhere under SSA Form}

\authorinfo
{Florent Bouchez}
{ENS-Lyon}
{}

\authorinfo
{Alain Darte}
{CNRS}
{}

\authorinfo
{Fabrice Rastello}
{INRIA}
{}

\authorinfo
{}
  {Université de Lyon, LIP, ENS Lyon, UCBL, CNRS, INRIA, France}
  {firstname.lastname@ens-lyon.fr}

\subtitle{Research Report n\textsuperscript{o} RR2007-42}



\maketitle

\begin{abstract}
  Compilation for embedded processors can be either aggressive (time
  consuming cross-compilation) or just in time (embedded and usually
  dynamic). The heuristics used in dynamic compilation are highly
  constrained by limited resources, time and memory in particular.
  Recent results on the SSA form open promising directions for the
  design of new register allocation heuristics for embedded systems
  and especially for embedded compilation. In particular, heuristics
  based on tree scan with two separated phases --- one for spilling,
  then one for coloring/coalescing --- seem good candidates for
  designing memory-friendly, fast, and competitive register
  allocators. Still, also because of the side effect on power
  consumption, the minimization of loads and stores overhead (spilling
  problem) is an important issue. This paper provides an exhaustive
  study of the complexity of the ``spill everywhere'' problem in the
  context of the SSA form.  Unfortunately, conversely to our initial
  hopes, many of the questions we raised lead to NP-completeness
  results.  We identify some polynomial cases but that are impractical
  in JIT context. Nevertheless, they can give hints to simplify
  formulations for the design of aggressive allocators.
\end{abstract}




\section{Introduction}

\def\omeg1{\Omega'\leq\Omega - 1}
\def\omeglarge{\Omega'\leq\Omega - k}
\def\omegr{\Omega'\leq r}
\def\omegfew{\Omega'\leq k}
\def\maxlive{\textrm{Maxlive}}

Register allocation is one of the most studied problems in
compilation.  Its goal is to map the temporary variables used in a
program to either machine registers or main memory locations.  The
complexity of register allocation for a fixed schedule comes from two
main optimizations, {\it spilling} and {\it coalescing}. Spilling
decides which variables should be stored in memory to make possible
register assignment (the mapping of other variables to registers)
 while minimizing the overhead of stores and loads.  Register
coalescing aims at minimizing the overhead of moves between registers.

Compilation for embedded processors is either aggressive or just in
time (JIT). Aggressive compilation is allowed to use a long compile
time to find better solutions. Indeed, the program is usually
cross-compiled, then loaded in permanent memory ({\sc rom}, flash,
etc.), and shipped with the product.  Hence the compilation time is
not the main issue as compilation happens only once.  Furthermore,
especially for embedded systems, code size and energy consumption
usually have a critical impact on the cost and the quality of the
final product. Just-in-time compilation is the compilation of code on
the fly on the target processor.  Currently the most prominent
languages are CLI and Java. The code can be uploaded or sold
separately on a flash memory, then compilation can be performed at
load time or even dynamically during execution. The heuristics used,
constrained by time and limited memory, are far from being aggressive.
In this context there is trade-off between resource usage for
compilation and quality of the resulting code.

\subsection{SSA Properties}
The static single assignment (SSA) form is an intermediate
representation with very interesting properties. A code is in SSA form
when every scalar variable has only one textual definition in the
program code. Most compilers use a particular SSA form, the strict SSA
form, with the additional so-called dominance property: given a use of
a variable, the definition occurs before any uses on any path going
from the beginning of the program (the root) to a use. One of the
useful properties of such a form is that the dominance graph is a tree
and the live ranges of the variables (delimited by the definition and
the uses of a variable) can be viewed as subtrees of this dominance
tree.  A well-known result of graph theory states that the
intersection graph of subtrees of a tree is chordal (see details
in~\cite[p.~92]{Golumbic}).  Since coloring a chordal graph is easy
using a greedy algorithm, it has the consequence for register
allocation that the ``assignment problem''~\cite[p.~622]{Cooper}
(mapping of variables to registers with no additional spill) is also
easy.

The fact that the interference graph of a strict SSA code is chordal,
and therefore easy to color, leads to promising directions for the
design of new register allocation heuristics.

\subsection{Recent Developments in Register Allocation}
Spilling and coalescing are correlated problems that are, in classical
approaches, done in the same framework. Even if ``splitting'', i.e.,
adding register-to-register moves, is sometimes considered in such a
framework, it is very hard to control the interplay between spilling
and splitting/coalescing. The properties of SSA form has led to new
approaches where spilling and coalescing are treated separately: the
first phase of spilling decides which values are spilled and where, so
as to get a code with $\maxlive\leq k$ where $\maxlive$ is the maximal
number of variables simultaneously live and $k$ is the number of
available registers.  The second phase of coloring (assignment), maps
variables to registers with no additional spill. When possible, it
also removes move instructions, also called shuffle code
in~\cite{LuehGross00}, due to coalescing. This is the approach
advocated by Appel and George~\cite{AppelGeorge01} and, more recently,
in~\cite{Brisk,HackGrundGoos06,BouchezDGR05,BouchezDGR06b}. The
interest of this approach for embedded systems is twofold.
\begin{enumerate}
\item Because power consumption has to be minimized, it is very
  important to optimize memory transfers and thus design heuristics
  that spill less. This new approach allows to design much more
  aggressive spilling algorithms for aggressive compilers.
\item For JIT compilation, this approach allows to design very fast
  spilling heuristics. In a graph coloring
  approach~\cite{CooperDasgupta06}, the spilling decision is
  subordinate to coloring. On the other hand, when the spilling phase
  is decoupled from the coloring/coalescing phase, i.e., when one
  considers better to avoid spilling at the price of
  register-to-register moves, then testing if spilling is required
  simply relies on checking that the number of simultaneous live
  variables (register pressure) is lower than $k$. This simple test
  can be performed directly on the control flow graph and the
  construction of an interference graph can thus be avoided. This
  point is especially interesting for JIT compilation since building
  an interference graph is not only time
  consuming~\cite{CooperDasgupta06}, but also memory
  consuming~\cite{Budimlic02}.
\end{enumerate}

The second advantage of the dominance property under SSA form is that the
coloring can be performed greedily on the control flow graph. The
principle for coloring a program under SSA form can be seen as a
generalization of linear scan.
\subparagraph{Linear scan:}
In a linear scan algorithm, the program is mapped to a linear
sequence. On this sequence, the live range of a variable is an union
of intervals with gaps in between.  The sequence is scanned from top
to bottom and, when an interval is reached, it is given an available
color, i.e., not already used at this point.  In Poletto and
Sarkar's approach~\cite{PolettoSarkar99}, each variable is
pessimistically represented by a unique interval that contains all the
effective intervals (the gaps are ``filled'').  It has the negative
effect of overestimating the register pressure between real intervals
but it ensures that all intervals of the same variable are assigned
the same register. In some way, Poletto and Sarkar's algorithm
provides a ``color everywhere'' allocation, i.e., it does not perform any
live-range splitting. Allowing the assignment of different colors for
a given variable requires shuffle code~\cite{TraubHS98,WimmerM05} to
be inserted afterwards to repair inconsistencies.  Such a repairing
phase requires additional data-flow analysis that might be too costly
in JIT context.

\subparagraph{Tree scan:}  Coloring a program under SSA can be seen as
a tree scan: the program is mapped on the dominance tree, live ranges
are subtrees. The dominance tree is scanned from root to leaves and
when an interval is reached it is given an available color. Here the
liveness is accurate and there is no need for gap filling or
additional live range splitting. Replacing $\phi$-functions by shuffle
code does not require any global analysis. In other words, tree scan
is a generalization of linear scan.

\subsection{Spill Everywhere}
As already mentioned, the dominance property of SSA form suggests
promising directions for the design of new register allocation
heuristics especially for JIT compilation on embedded systems. The
motivation of our study was driven by the hope of designing both fast
and efficient register allocation based on SSA form. Notice that
answering whether spilling is necessary or not is easy --- even if
there can be some subtleties~\cite{BouchezDGR06b} --- while minimizing
the amount of load and store instructions is the real issue. In other
words, if the search space is now cleanly delimited, the objective
function that corresponds to minimizing the spill cost has still some
open issues. So the question is: Is it easier to solve the spilling
problem under SSA? In particular is the spill everywhere problem
simple under SSA form?

The spilling problem can be considered at different granularity
levels: the highest, so called spill everywhere, corresponds to
considering the live range of each variable entirely. A spilled
variable will then lead to a store after the definition and a load
before each use.  The finer granularity, so called load-store
optimization, corresponds to optimize each load and store separately.
The latter problem, also known as paging with write back, is
NP-complete~\cite{Liberatore00} on a basic block even under SSA form.
The former problem is much simpler, and a well-known polynomial
instance~\cite{belady66} exists under SSA form on a basic block.  To develop new spilling heuristics, studying the complexity of
spilling everywhere is very important for the design of both
aggressive and JIT register allocators.
\begin{enumerate}
\item First, the complexity of the load-store optimization problem
  comes from the asymmetry between loads and
  stores~\cite{Liberatore00}. The main difference between the
  load-store optimization problem and the spill everywhere problem
  comes from this asymmetry.  We have measured that, in practice, most
  SSA variables have only one or two uses. So, it is natural to wonder
  whether this singularity makes the load-store optimization problem
  simpler or not.  The extreme case with only one use per variable is
  equivalent to the spill everywhere problem. More generally, even in
  the context of a traditional compiler, the spill everywhere problem
  can be seen as an oracle for the load-store optimization problem to
  answer whether a variable should be stored or not. In the context of
  aggressive compilation~\cite{Hack06ssaopt,lcpcgrothoff}, a way to
  decrease the complexity is to restore the
symmetry between loads and stores as done
in~\cite{AppelGeorge01}\footnote{In this formulation, a variable
  might be either in memory location or in a register, but cannot reside in
  both.}.
\item Second, spill everywhere is a good candidate for designing
  simple and fast heuristics for JIT compilation on embedded systems.
  Again, in this context, the complexity and the footprint of the
  compiler is an issue. Spilling only parts of the live ranges, as
  opposed to spilling everywhere, leads to irregular live range
  splitting and the insertion of shuffle code to repair
  inconsistencies, in addition to maintaining liveness information for
  coalescing purpose. All of this is probably too costly for some
  embedded compilers.
\end{enumerate}
Studying the complexity of the spill everywhere problem in the context
of SSA form is thus important to guide the design of both aggressive
and JIT register allocation algorithms. This the goal of this paper.
To our knowledge this is the first exhaustive study of this problem in
the literature.

\subsection{Overview of the paper}

The rest of paper is organized as follows. For our study, we
considered different variants of the spilling problem.
Section~\ref{sec_terminology} provides the terminology and notation
that describe the different cases we considered.
Section~\ref{sec_no_holes} considers the simplified spill model
where a spilled variable frees a register for its whole live range; we
provide an exhaustive study of its complexity under SSA form.
Section~\ref{sec_with_holes} deals with the problem where a spilled
variable might still need to reside in a register at its points of
definition and uses. Here, the study is restricted to basic blocks as
it is already NP-complete for this simple case.
Section~\ref{sec_conclusion} summaries our results and concludes.

\section{Terminology and Notation}

\label{sec_terminology}
\paragraph{Context:}
For the purpose of our study, we consider different configurations
depending whether live ranges are {\em restricted to a basic block or
  not}. Indeed, on a basic block, the interference graph is an
interval graph, while for a general control flow graph, under strict
SSA form, it is chordal.
We also consider whether the use of an evicted variable in an
instruction requires a register or not. If not, spilling a variable
corresponds to decreasing by one the register pressure on every points
of the corresponding live range. Otherwise, spilling a variable does
not decrease the register pressure on program points that use it: in
that case, instead of having the effect of removing the entire live
range, spilling a variable corresponds to removing a version of the
live range with ``holes'' at the use and definition points. We denote
those two problems respectively as {\em without holes} or {\em with
  holes}.  Finally, we distinguish the cases where the cost of
spilling is the same for all variables or not. We denote those two
problems respectively as {\em unweighted} (denoted by $w(v)=1$ for all
$v$) or {\em weighted} (denoted by $w\neq 1 $).

\paragraph{Decreasing \maxlive:}
As mentioned earlier the goal of the spilling problem is simply to
lower the register pressure at every program point, while the
corresponding optimization problem is to minimize the spilling
cost. At a given program point, the register pressure is the number of
variables alive there. The maximum over all program
points, usually named $\maxlive$, will be denoted by $\Omega$ here. Let us
denote by $r$ the number of available registers. Hence formally, the
goal is to decrease $\Omega$ by spilling some variables. If we denote
by~$\Omega'$ the register pressure after this spilling phase, we 
distinguished the following four problems: $\omeg1$, $\omeglarge$ where
$k$ is a constant, $\omegfew$ where $k$ is a constant, and
the general problem $\omegr$ where there is no constraint on the
number of registers $r$.

\paragraph{A graph problem:}
The spill everywhere problem without holes can be expressed as a node
deletion problem~\cite{Yannakakis78}. The general node deletion
problem can be stated as follows: ``Given a graph or digraph $G$ find
a set of nodes of minimum cardinal, whose deletion results in a
subgraph or subdigraph satisfying the property~$\pi$.'' Hence, the
results of the first section have a domain of application not only on
register allocation but also on graph theory. For this reason, we
formalize them using graphs (properties of the interference graphs)
instead of programs (register pressure on the control flow graph)
while the algorithmic behind is actually based on the control flow
graph representation.

\paragraph{Perfect graphs:}
Perfect graphs~\cite{Golumbic} have some interesting properties for
register allocation. In particular, they can be colored in polynomial
time, which suggests that we can design heuristics for spilling or
coalescing in order to change the interference graph into a perfect
graph. For a graph~$G$, the maximal size of a complete subgraph, i.e.,
a clique, is the {\it clique number} $\omega(G)$. The minimum number
of colors needed to color $G$ is the {\it chromatic number}~$\chi(G)$.
Of course, $\omega(G) \leq \chi(G)$ because vertices of a clique must
have different colors. A graph $G$ is perfect if each induced
subgraph~$G'$ of~$G$ (including $G$ itself) is such that
$\chi(G')=\omega(G')$. A {\it chordal} graph is a perfect graph;
it is the intersection graph
of subtrees of a tree: to each subtree corresponds a vertex, and
there is an edge between two vertices if the corresponding subtrees
intersect. A well-known subclass of
chordal graphs is the class of interval graphs, which are 
intersection graphs of subsequences of a sequence. 

\section{Spill Everywhere without Holes}
\label{sec_no_holes}

It is well-known that, on a basic block, the unweighted spill
everywhere problem without holes is polynomial: this is the greedy
furthest use algorithm described by Belady~\cite{belady66}.  It is
less known that the weighted version of this problem, which cannot be
solved using this last technique, is also
polynomial~\cite{YannakakisGavril87,Liberatore00}: the interference
graph is an intersection graph for which the incidence matrix is
totally unimodular and the integer linear programming (ILP)
formulation can be solved in polynomial time. This property holds also
for a path graph, which is a class of intersection graphs between
interval graphs and chordal graphs. We recall these results here for
completeness. We also recalled earlier that, under SSA form, once the
register pressure has been lowered to $r$ at every program point, the
coloring ``everywhere'' problem (each variable is assigned to a
\emph{unique} register) is polynomial.

The natural question raised by these remarks is whether the spill
everywhere problem without holes is polynomial or not. In other words,
does the SSA form make this problem simpler? The answer is no. A graph
theory result of Gavril and Yannakakis~\cite{YannakakisGavril87} shows
it is NP-complete, even in its unweighted version: for an arbitrarily
large number of registers $r$, a program with $\Omega$ arbitrarily
larger than~$r$, spilling everywhere a minimum number of variables
such that $\Omega'$ is at most $r$ is NP-complete. The
main result of this section shows more: this problem remains
NP-complete even if one requires only $\omeg1$.  The practical
implication of this result is that for a heuristic that would lower
$\Omega$ one by one iteratively, even the optimization of each
separate step is an NP-complete
problem.\footnote{Note that providing an optimal solution for each
  intermediate step (going from $\Omega$ to $\Omega-1$, then from
  $\Omega-1$ to $\Omega-2$, and so on, until $\Omega'=r$) does not
  always give an optimal solution for the problem of going from
  $\Omega$ to $r$.
}

Table~\ref{tab.spillnoholes} summarizes the complexity results of
spilling everywhere (without holes). We now recall classical results
and prove new more accurate results. Let us start with the decision
problem related to the most general case of spill everywhere without
holes.

\def\noteontablearrows{
{\small Note:
weaker results have arrows pointed to the proof subsuming them.
}
}

\begin{table*}[Htp]
  \begin{center}
    \begin{tabular}{|c|c|c|c|c|}
      \hline
      & weighted & $\omegfew$ & $\omegr$ & $\omeg1$ \\
      \hline
      Chordal graph & no & \P\ $\downarrow$ & \NP\ 
      \ifx\usetikz\undefined{
      $\rightarrow$}
      \else{
      \tikz { \node at (0,0) {} ; \draw[->,yshift=0.6pt] (0,0) -- +(1.1,0) { }; }
      }
      \fi
      & \NP\ 3-exact 
      cover \\
      {\it = general SSA case}      & yes & \P\ dynamic prog. & \NP\
      \ifx\usetikz\undefined{
      $\nearrow$ 
      }\else{
      \tikz { \draw[->] (0,0) -- +(1.1,0.18) { }; }
      }\fi
      & \NP\ $\uparrow$ \\
      \hline
      Interval graph & no & \P\ $\uparrow$ & \P\ greedy (furthest use) & \P\ 
      $\downarrow$ \\
      {\it = basic block} & yes & \P\ $\uparrow$ & \P\ ILP & \P\ dynamic prog. \\
      \hline
    \end{tabular}

    \noteontablearrows
  \end{center}
  \caption{Spill everywhere without holes.}
  \label{tab.spillnoholes}
\end{table*}

\PROBLEM{Spill everywhere}{A perfect graph $G=(V,E)$ with clique
  number $\Omega=\omega(G)$, a weight $w(v)>0$ for each vertex, an
  integer $r$, an integer~$K$.}{Can we remove the vertices in
  $V_S\subseteq V$ from $G$ with overall weight $\sum_{v\in V_s}
  w(v)\leq K$ such that the clique number $\Omega'$ of the induced
  subgraph~$G'$ is at most $r$?}

\begin{theorem}[Furthest First]
\label{thm.belady}
The spill everywhere problem for an interval graph 
is polynomially solvable, with a
greedy algorithm, 
if $w(v)=1$ for all $v$ even if $r$ is not fixed.
\end{theorem}
The algorithm behind this theorem is the well-known furthest use
strategy described by Belady in~\cite{belady66}. This strategy is very
interesting for designing spilling heuristics on the dominance tree
(see for example~\cite{HGG05:RR}). We give here a constructive proof
for completeness.

\begin{proof}
  An interval graph is the intersection graph of a family of
  sub-sequences of a (graph) chain. For convenience, we denote the
  chain as $B$, vertices of $B$ are called points, and sub-sequences
  of~$B$ are called variables. Consecutive points are denoted by
  $p_1$, \ldots, $p_m$, and the set of variables is denoted by $V$.
  Once variables are removed (spilled), the remaining set of variables
  $V'$ is called an allocation. An allocation is said to fit $B$ if,
  for each point $p$ of $B$, the number of remaining variables
  intersecting $p$ is at most~$r$. The goal is to remove a minimum
  number of variables such that the remaining allocation fits $B$.
The greedy algorithm can be described as follows:
\begin{description}
\item[Step 0 (init)] Let $V'_0=V$ and $i=1$;
\item[Step 1 (find first)] Let $p(i)$ be the first point from the
  beginning of the chain such that more than $r$ remaining variables,
  i.e., in $V'_{i-1}$, intersect $p(i)$; 
\item[Step 2 (remove furthest)] Select a variable $v_i$ that
  intersects $p$ and ends the furthest and remove it, i.e., let $V'_i=V'_{i-1}\backslash \{v_i\}$;
\item[Step 3 (iterate)] If $V'_i$ fits $B$, stop, otherwise increment
  $i$ by $1$ and go to Step~1.
\end{description}

Let us prove that the solution obtained by the greedy algorithm is
optimal. Consider an optimal solution $S$ (described by a set
$V_{S}$ of spilled variables) such that $V_{S}$ contains the
maximum number of variables $v_i$ selected by the greedy algorithm.
Suppose that $S$ does not spill all of them and denote by $v_{i_0}$
the variable with smallest index such that $v_{i_0} \notin V_{S}$.
By definition of $p_{i_0}$ in the greedy algorithm, there are at least
$r+1$ variables not in $\{v_1, \ldots, v_{i_0-1}\}$ intersecting
$p(i_0)$. As $S$ is a solution, there is a variable $v$ in $V_{S}$
(thus $v\neq v_{i_0}$) that intersects $p(i_0)$. We claim that
spilling $W = V_{S} \cup \{v_{i_0}\} \setminus \{v\}$, i.e.,
spilling $v_{i_0}$ instead of $v$, is a solution too. Indeed, for all
points before $p(i_0)$ (excluded), the number of variables in
$V'_{i_0-1}=V\setminus\{v_1, \ldots, v_{i_0-1}\}$ is at most $r$.
Since $\{v_1, \ldots, v_{i_0}\} \subseteq W$, this is true for
$V\setminus W$ too.  Furthermore, each point $p$ after $p(i_0)$
(included), intersected by $v$, is also intersected by $v_{i_0}$by
definition of $v_{i_0}$.  Thus, as $p$ is intersected by at most $r$
variables in $V\setminus V_{S}$, the same is true for $V \setminus
W$. Finally, this solution spills more variables $v_i$ than $S$, which
is not possible by definition of $S$. Thus $V_S$ contains all variables
$v_i$ and, by optimality, only those. This proves that the greedy
algorithm gives an optimal solution.
\end{proof}
 
\begin{theorem}[poly. ILP]
\label{thm.ilp}
The spill everywhere problem for an interval graph is
polynomially solvable even
if $w\neq 1$ and $r$ is not fixed.
\end{theorem}
This result was pointed out by Gavril
and Yannakakis in~\cite{YannakakisGavril87} and used in a slightly
different context by Farach-Colton and Liberatore~\cite{Liberatore00}.
The idea is to formulate the problem using ILP and to remark that the
matrix defining the constraints is totally unimodular.  For the
sake of completeness, we provide the formulation here.

\begin{proof}
  We use the same notations as for Theorem~\ref{thm.belady} except
  that, now, $v_1$, \ldots, $v_n$ denote all variables and not only
  those selected by the greedy algorithm. Let $w_i$ be the cost of
  removing (spilling) variable~$v_i$.  We define the clique matrix as
  the matrix ${\cal C}=\left(c_{p,v}\right)$ where $c_{p,v}=1$ if $v$
  intersects the point $p$ and $c_{p,v}=0$ otherwise. Such a matrix is
  called the incidence matrix of the interval hyper-graph and is
  totally unimodular~\cite{berge}. The optimization problem can be
  solved using the following integer linear program, where $\vec{x}$
  is a vector with components $(x_i)_{1\leq i \leq n}$, $\vec{w}$ is a
  vector with components $(w_i)_{1\leq i \leq n}$, $\vec{r}$ is a vector
  whose components are all equal to $r$, and vector inequalities are
  to be understood component-wise:
  \[\max\left\{ \vec{w}.\vec{x}~|~{\cal C}\vec{x} \leq \vec{r},\, \vec{0}
    \leq \vec{x} \leq \vec{1}\right\}\] Of course, $x_i=0$ means that
  $v_i$ should be removed while $x_i=1$ means it should be kept. The matrix
  of the system is ${\cal C}$ with some additional identity matrices,
  which keeps the total unimodularity.
\end{proof}

The next two theorems are from Yannakakis and
Gavril~\cite{YannakakisGavril87}. 
\begin{theorem}[Yannakakis]
\label{thm.yannakakis_np} 
The spill everywhere problem is NP-complete for a chordal graph even
if $w(v)=1$ for each
$v\in V$.
\end{theorem}

Another important result of~\cite{YannakakisGavril87} is that the
spill everywhere problem is polynomially solvable when $r$ is fixed.
Of course, there is a power of $r$ in the complexity of their
algorithm, but it means that if $r$ is small, the problem is simpler.
Because of this, we call the problem when $r$ is fixed ``spill
everywhere {\em with few registers}''.

\PROBLEM{Spill everywhere with few registers ($k$)}{A perfect graph $G=(V,E)$ with clique
number $\Omega$, a weight $w(v)>0$ for each vertex, an
integer $K$, $r=k$ is fixed.}{Can we remove vertices $V_S\subseteq V$ from $G$ with overall weight
$\sum_{v\in V_s} w(v)\leq K$ such that the induced subgraph $G'$ has
  clique number $\Omega'\leq r$?}

\begin{theorem}[Dynamic programming on non-spilled variables]
\label{thm.yannakakis_poly}
The spill everywhere problem with few registers is polynomially solvable if~$G$ is chordal even if $w\neq 1$.
\end{theorem}

When we proved our results, we were actually not aware of Gavril and
Yannakakis paper. Since Theorem~\ref{thm.yannakakis_poly} is very
intuitive, we logically ended with the same kind of construction.  For
completeness, we provide it here, with our own notations.  This
proof is constructive and the algorithm (dynamic programming on
program points) is based on a tree traversal.  It performs
$\cplx(m\Omega^k)$ steps of dynamic programming, where $m$ is the
number of program points.

\begin{proof} 
  A chordal graph is the intersection graph of a family $V$ of
  subtrees of a tree $T$ (Thm~4.8~\cite{Golumbic}). We call {\it
    points} the vertices of the tree $T$ and, to distinguish the
  maximal subtrees $T_p$ rooted at each given point $p$ from the
  subtrees of the family $V$, we call the latter {\it variables}.
  Given a point $p$ and a set $W\subseteq V$ of variables, let
  $W(p)$ be the set of variables $v\in W$ intersecting~$p$, i.e., such
  that $p$ belongs to the subtree $v$. If $|W(p)| \leq r$, we say
  that~$W$ fits $p$ and that $W(p)$ is a fitting set for $p$. We say
  that $W$ fits a set of points if it fits each of these points. A
  solution to the spill everywhere problem with $r$ registers is thus
  a subset $W$ of $V$ such that $W$ fits $T$. It is an optimal
  solution if $\sum_{v\in W} w(v)$ is maximal.  With these notations,
  $W$ corresponds to $V-V_S$ in the spill everywhere problem
  formulation, and maximizing the cost of~$W$ is equivalent to
  minimizing the weight of~$V_S$.

  Given a subset of variables $W$, we consider its {\it restriction},
  denoted by $W_p$, to a subtree $T_p$: it is defined as the set of
  variables $v \in W$ that have a non-empty intersection with $T_p$.
  Note that if~$W$ fits $T$, then its restriction $W_p$ to a
  subtree~$T_p$ fits $T_p$.  Furthermore, if~$p_1$ and $p_2$ are
  children of $p$ in $T$ then, because of the tree structure, all
  variables that belong to both $W_{p_1}$ and $W_{p_2}$ intersect~$p$,
  and all variables in $W_{p_i}$ intersecting $p$ intersect also
  $p_i$, i.e., $W_{p_i}(p) =W_p(p_i)$.  These remarks ensure the
  following.  Let~$W$ be a fitting set for $T_p$
  and let $W'$ be a fitting set for $T_{p_i}$ such that $W'_{p_i}(p) =
  W_{p_i}(p)$ (i.e., they coincide between $p$ and $p_i$). Then, replacing
  $W_{p_i}$ by $W'_{p_i}$ in~$W$ leads to another fitting set of
  $T_p$.
  This is the key to get an optimal solution thanks to
  dynamic programming.

  The final proof is an induction on the points $p$ of $T$ --- from
  the leaves to the root --- and on the fitting sets of those points
  $F_p\in {\cal F}_p=\left\{W \subseteq V(p); |W|\leq r\right\}$. Let
  us denote by $W_{max}(p,F_{p})$ a subset $W$ of $V$ that contains
  only variables intersecting $T_p$, such that $W(p) = F_p$, and with
  maximal cost. It can be built recursively as follows. For each child
  $p_i$ of~$p$, consider all possible fitting sets~$F_{p_i}$ that
  match $F_p$, i.e., such that $F_{p_i}\cap V(p)=F_p\cap V(p_i)$ and
  pick the solution such that $W_{max}(p_i, F_{p_i})$ is maximal.
  From these selected subsets, one for each $p_i$, $W_{max}(p, F_p)$
  can be defined. This construction is done for each $F_p \in {\cal
    F}_p$. As there are at most $V(p)^k\leq \Omega^k$ such fitting
  sets for $p$, these successive locally optimal solutions can be
  built in polynomial time.
\end{proof}

We now address the following problem, which is a particular case of
the more general spill everywhere problem.

\PROBLEM{Incremental spill everywhere}{A perfect graph $G=(V,E)$ with
  clique number $\Omega=\omega(G)$, a weight $w(v)>0$ for each vertex,
  an integer $K$.}{Can we remove vertices $V_S\subseteq V$ from $G$ with
  overall weight $\sum_{v\in V_s} w(v)\leq K$ such that the induced
  subgraph $G'$ has clique number $\Omega'\leq \Omega-1$?}

The following theorem can be seen as a particular case of
Theorem~\ref{thm.ilp}. The proof is interesting since it provides an
alternative solution to the ILP formulation for this simpler case.

\begin{theorem}[Dynamic programming on spilled variables]
\label{thm.dynamic}
If $G$ is an interval graph, the incremental spill everywhere problem
is polynomially solvable, even if $w\neq 1$.
\end{theorem}

\begin{proof}
  Let $B=\{p_1,\ldots, p_m\}$ be a linear sequence of points,
  $p_i < p_j$ if $i<j$, and $V=\{v_1, \ldots, v_n\}$ be a set of
  weighted variables, where each variable $v_i$ corresponds to an
  interval $[s(v_i),e(v_i)]$. We assume that the variables are sorted
  by increasing starts, i.e., $s(v_i) \leq s(v_j)$ if $i<j$. Without
  loss of generality, the problem can be restricted to the case where
  any point~$p$ belongs to exactly $\Omega$ variables (any other point
  can be deleted from the instance). So for each point, one needs to
  spill at least one of the intersecting variables. What we seek is
  thus a minimum weighted cover of $B$ by the variables of~$V$, which
  can be done thanks to dynamic programming as follows.

  Let $W(p_i)$ be the minimum cost of a cover of $p_1$, \ldots,
  $p_i$.
  Knowing all $W(p_{j<i})$,
it is possible to compute $W(p_i)$.
Indeed, at~$p_i$, one must choose a variable $v \in V(p_i)$, i.e.,
intersecting the point $p_i$. As~$v$ already covers the interval
between its start $s(v)$ and $p_i$, we get:
  \[
  W(p_i) = \min_{v\in V(p_i)} (w(v) + W(\mbox{pred}[s(v)])) \mbox{ where
    } \mbox{pred}[p_i]=p_{i-1} 
  \]
  with the convention $W(p) = 0$ for $p<p_1$.  $W(p_m)$ is the minimum
  cost of an incremental spilling over the whole basic block~$B$.  The
  set~$V(p_i)$ can be computed from $V(p_{i-1})$ in $\cplx(\Omega)$ operations
  because the variables are sorted by increasing starts. The overall
  complexity is thus $\cplx(\Omega m)$.
%
%
\end{proof}

\begin{theorem}[From 3-exact cover]
\label{thm.3exact}
The incremental spill everywhere problem is NP-complete for a chordal
graph even if
$w(v)=1$ for each $v\in V$.
\end{theorem}

\begin{proof}
  As for Theorem~\ref{thm.yannakakis_poly} we use the characterization
  of a chordal graph as an intersection graph of a family of subtrees
  of a tree. We use the same notations. The proof is a reduction from
  \textsf{Exact Cover by 3-Sets} (X3C)~\cite[Problem SP2]{gareyjohnson}: let
  $\mathcal{P}$ be a set of $3n$ elements $\{ p_1, p_2, \cdots,
  p_{3n}\}$, and $\mathcal{V} = \{ v_1, v_2, \cdots, v_m \}$ a set of
  subsets of $\mathcal{P}$ where each subset contains exactly three
  elements of $\mathcal{P}$. Does $\mathcal{V}$ contains an exact
  cover of $\mathcal{P}$, i.e., a sub-collection $\mathcal{S}\subseteq
  \mathcal{V}$ such that every element of $\mathcal{P}$ occurs in
  exactly one member of $\mathcal{S}$?

  Let us consider an instance of X3C and define the following family
  of subtrees of a tree: the main tree $T$ is of height 2 with one
  root point labeled $p_0$ and $3n$ leaves labeled $p_1, p_2, \cdots,
  p_{3n}$. For each $v_i=\{p_{\alpha},p_{\beta},p_{\gamma}\}$ there is
  a subtree (variable) made of the root $p_0$ and the tree points
  $p_{\alpha},p_{\beta},p_{\gamma}$.  The number of variables
  intersecting~$p_0$ is
  $m$, so 
  $\Omega=m$. Let us create as many additional variables as necessary
  (we call them non-labeled variables) so that the number of
  intersecting variables is exactly $\Omega$ for each point of $T$. In
  other words, for a leaf $p_j$ that belongs to $k$ subtrees $v_i$, we
  create $m-k$ subtrees, each containing only $p_j$.  Given this
  family of subtrees of a tree, consider the corresponding
  intersection graph (which is chordal).  We now show that this
  instance of X3C has a solution if and only if it is possible to
  remove (spill) at most $n=K$ variables such that, for each point
  $p$, the number of remaining intersecting variables is at most
  $\Omega-1$.  Notice that the reduction is polynomial: the whole
  number of variables is not larger than $3n\times m$.

  Suppose that there is a solution to the incremental spill
  everywhere problem and let $V_S$ be the set of removed variables
  with $|V_S|\leq n$.  There is no non-labeled variable in~$V_S$
  because $\Omega$ must be decreased in the $3n$ leaves and only a
  labeled variable goes over three leaves.  Hence $V_S$ contains only
  labeled variables, $|V_S| = n$, and the corresponding set of subsets
  $\mathcal{S}$ is a covering of~$\cal{P}$.  Conversely, suppose that
  the X3C instance has a solution~$\mathcal{S}$ and let~$V_S$ be the
  set of corresponding subtrees. Since $\mathcal{S}$ is a covering of
  $\mathcal{P}$, $|\mathcal{S}|=n$ and there is exactly one
  intersecting set in $V_S$ for each leaf. So the number of remaining
  intersecting variables is $\Omega-1$ for each leaf. As for the root
  $p_0$, all variables intersect it, so there is at least one
  (labeled) variable removed and the number of remaining intersecting
  variables is at most $\Omega-1$. In other words, $V_S$ is a
  solution, with $|V_S|\leq n$, to the incremental spill
  everywhere problem.

  This proves that the incremental spill everywhere problem is
  NP-complete (the fact it belongs to NP is straightforward).
\end{proof}

The comparison between this last theorem and
Theorem~\ref{thm.yannakakis_poly} is very interesting. Indeed, our
first (false) intuition was that choosing which variables to remove so
as to go from $\Omega$ to $\Omega-k$ was exactly the symmetric of
choosing which variables to keep so as to get down to $k$. At first
sight, it seemed that dynamic programming could be used, as for
Theorem~\ref{thm.yannakakis_poly}, to solve the incremental spill
everywhere problem. For interval graphs, both problems can indeed be
solved with dynamic programming as we previously showed. The
incremental approach would have then provided a heuristic for the main
spill everywhere problem, as an alternative to an exact solution as
in~\cite{AppelGeorge01}, which is too expensive when $r$ is large.
Unfortunately, Theorem~\ref{thm.3exact} contradicts this intuition. In
fact, the two problems are not perfectly symmetric: to make the graph
$k$-colorable, the number of kept variables live at any point should
be \emph{at most} $k$ while to make a graph $\Omega-k$ colorable, the
number of removed variables live at any point must be \emph{at least}
$k$, as for the point $p_0$ in the proof of Theorem~\ref{thm.3exact}.
This is where the combinatorial complexity comes from.

\section{Spill Everywhere with Holes on a Basic Block}
\label{sec_with_holes}

The previous section dealt with the spill everywhere problem without
holes. To summarize, this problem is polynomial for a basic block even in
its weighted version whereas, most of the time, it is NP-complete for a
general control flow graph under SSA form. As mentioned earlier,
the model without holes does not reflect the reality of most
architectures. The goal of this section is to tackle the problem of
spill everywhere with holes on a basic block.

Where do the holes come from? For an architecture where operations are
allowed only between registers, whenever a variable is spilled, one
needs to insert load instructions before the uses of this variable and
a store instruction after its definition. This means that new
variables appear, with very short live ranges but which nonetheless
need to be assigned to registers. In other words, when a variable is
spilled, the number of simultaneously alive variables decreases by one
at every point of the live range, \emph{except} where the variable is
defined or used.  Thus spilling everywhere a variable does not remove
the complete  interval, but only parts of it, since there
is still some tiny sub-intervals left.  This is why, for instance, in
Chaitin et al.  algorithm~\cite{Chaitin81}, the register allocation
must re-build the interference graph and iterate if some variables are
spilled.

\begin{figure*}[ht]
  \begin{center}
    \includegraphics[scale=0.65]{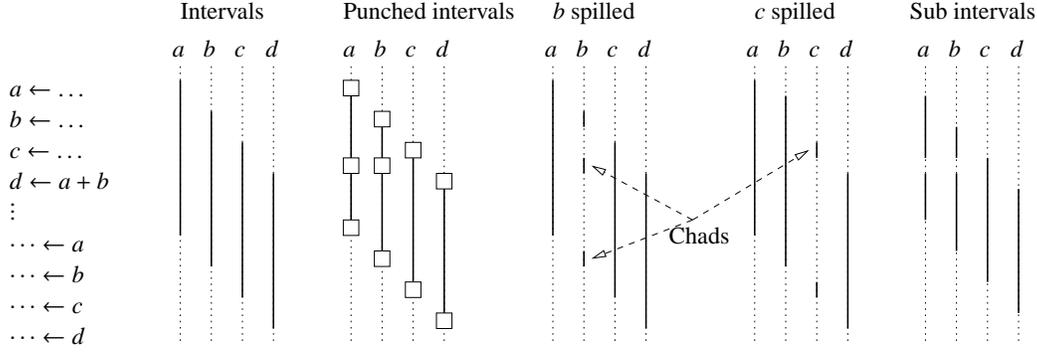}
  \end{center}
  \caption{Example of punched intervals.}
  \label{fig:trousex}
\end{figure*}

\paragraph{Holes and chads:}
The notion of holes can be formalized as follows. An SSA code on a
basic block, or \emph{linear SSA code}, is a pair $\mathcal{C}=(B,V)$
where $B = \{p_1, \ldots, p_m\}$ is a sequence of $m$ instructions; and
$V$ the set of variables which appear in those instructions.
An instruction {\em first} uses simultaneously some variables and {\em
  then} possibly defines some other new variables.  Each variable of
$V$ is defined at most once and, if it is not defined, it is live-in
for the sequence $B$.  Also, each variable either has a ``last use''
(last instruction which uses it) or is live-out for the sequence.  A
variable is represented by a simple interval of the sequence $B$,
starting at the middle of the instruction that defines it (or at the
beginning of~$B$ for a live-in), and ending at the middle of its last
use (or at the ending of~$B$ for a live-out).  {\em Spilling} a
variable $v\in V$ decreases by one the register pressure at each of
its points but not at its definition and uses points: the set of
points that is actually ``removed'' is the interval $v$ with holes
on it, so we call it a {\em punched interval}.  The remaining
points $c\in v$ which are not removed are called {\em chads}, as if,
when spilling the variable $v$, one first had punched the
corresponding interval, leaving small intervals in place.
See Figure~\ref{fig:trousex} for a graphical explanation.

\paragraph{Simultaneous holes:}
Also, we distinguish different cases depending on $h$, the
\emph{number of simultaneous holes}. This number correspond to the
maximum number of registers which can be used (arguments) by the same
instruction or defined by the same instruction.  For instance, $h=2$
in the following three operand addition
\verb+add %reg1, %reg2 => %reg3+.
Finally, for a given point~$p$ of~$B$, the set of variables live at
$p$ is denoted by $L(p)$. Its cardinal, the register pressure, is
denoted by $l(p)=|L(p)|$ and $\maxlive$, the maximum of $l(p)$ over
all points $p \in B$, is denoted by $\omega(\mathcal{C})$.  Once some
variables $V_S$ have been spilled, the induced code can be
characterized as follows. The set of spilled variables live at~$p$
is $L_S(p)=V_S\bigcap L(p)$; the set of non-spilled live variables is
$L'(p)=L(p)\backslash L_S(p)$. The new register pressure is denoted by
$l'(p)$. Notice that $L'(p)$ does not contain any chad, whereas of
course $l'(p)$ needs to take remaining chads into account. Hence
$l'(p)$ is not necessarily equal to $|L'(p)|$ but, more generally,
$|L'(p)|\leq l'(p)\leq |L'(p)|+h$.

\smallskip All previous notions can be generalized to a general SSA
program. The sequence $B$ (linear code) becomes a tree $T$ (dominance
tree) and punched intervals become punched subtrees. Now, the (general)
problem can be stated as follows.

\PROBLEME{Spill everywhere with holes}{A code $\mathcal{C}=(T,V)$ with
  $\maxlive$ $\Omega=\omega(\mathcal{C})$, a weight $w(v)>0$ for each
  variable, integers $r$ and $K$.}{Can we spill variables $V_S\subseteq
  V$ from $V$ with overall weight $\sum_{v\in V_s} w(v)\leq K$ such
  that the induced code $\mathcal{C}'$ has $\maxlive$ $\Omega'\leq
  r$?}{The spill everywhere {\em on a basic block} denotes the case
  where $T$ is a sequence $B$ (linear code). The spill everywhere {\em
    with few registers} ($k$) denotes the case where $r$ is fixed
  equal to $k$. The spill everywhere {\em with many registers} ($k$)
  denotes the case where $r$ is equal to $\Omega-k$. The {\em
    incremental} spill everywhere denotes the case where $r$ is equal
  to $\Omega-1$.}

As explained in~\cite{Liberatore00},
the hardness of load-store optimization comes from the fixed cost of
the store (once a variable is chosen to be evicted) while the number
of loads (number of times it is evicted) is not fixed.  Neglecting the
cost of the store would lead to a polynomial problem where each
sub-intervals of the punched interval could be considered
independently for spilling. But we feel that this approximation is not
satisfactory in practice because the mean number of uses for each
variable can be small. Indeed, we measured on our compiler tool-chain,
using small kernels representative of embedded applications, that most
spilled variables have at most two uses. Hence, minimizing the number
of spilled variables is nearly as important as minimizing the number
of unsatisfied uses. Consider for example a furthest-first-like
strategy on sub-intervals (see Figure~\ref{fig:trousex} for an
illustration of sub-intervals). To design such a heuristic, a spill
everywhere solution might be considered to drive decisions: between
several candidates that end the furthest, which one is the most
suitable to be evicted in the future? Unfortunately, as summarized by
Table~\ref{tab.spillwithholes}, most instances of spill everywhere
with holes are NP-complete for a basic block.

We start with a result similar
to Theorem~\ref{thm.yannakakis_poly}: even with holes, the spill
everywhere problem with few registers is polynomial.


\begin{table*}[!ht]
\begin{center}
\begin{tabular}{|c|c|c|c|c|c|}
\hline
  & weighted & $\omegfew$ & $\omegr$ & $\omeglarge$ & $\omeg1$ \\
\hline
$h=1$ & no & \P\ $\downarrow$ & ? & \P\ $\downarrow$ & \P\ $\downarrow$ \\
      & yes & \P\ $\downarrow$ & \NP\ stable set & \P\ $\downarrow$ & \P\ $\downarrow$ \\
\hline
$h\geq 2$ & no & \P\ $\downarrow$ & \NP\ stable set & \P\ $\downarrow$ & \P\ $\downarrow$ \\
      & yes & \P\ $\downarrow$ & \NP\ $\uparrow$ & \P\ dynamic prog. & 
      \ifx\usetikz\undefined{
      $\leftarrow$
      }\else{
\tikz { \node at (0,0) {} ; \draw[<-,yshift=0.6pt] (0,0) -- +(1.0,0) { }; }
}\fi
      \P\
\\
\hline
$h$ not bounded & no & \P\ $\downarrow$ & \NP\ 
      \ifx\usetikz\undefined{
$\rightarrow$
}\else{
\tikz { \node at (0,0) {} ; \draw[->,yshift=0.6pt] (0,0) -- +(1.1,0) { }; }
}\fi
& \NP\ 
      \ifx\usetikz\undefined{
$\rightarrow$
}\else{
\tikz { \node at (0,0) {} ; \draw[->,yshift=0.6pt] (0,0) -- +(1.1,0) { }; }
}\fi
& \NP\ set cover  \\
 & yes & \P\ dynamic prog. & \NP\ $\uparrow$ &  \NP\ $\uparrow$ & \NP\ $\uparrow$  \\
\hline
\end{tabular}\\
\noteontablearrows
\end{center}
\caption{Spill on interval graphs with holes.}
\label{tab.spillwithholes}
\end{table*}

\begin{theorem}[Dynamic programming on non-spilled variables]
\label{thm.dynamic_holes}
The spill everywh\-ere problem with holes and few registers is
polynomially solvable even if
$w\neq 1$. 
\end{theorem}
\begin{proof}
  The proof is similar to the proof of
  Theorem~\ref{thm.yannakakis_poly}. The only point is to adapt the
  notations to take chads into account. The word ``removed'' has to be
  replaced by ``spill'' since variables are not removed entirely.
  Furthermore, the definition of ``fitting set'' needs to be modified. A
  set $F_p$ of variables is a fitting set for $p$ if, when all
  variables not in $F_p$ are spilled, the new register pressure
  $l'(p)$ is at most $r$. In other words, the set of fitting sets
  becomes ${\cal F}_p=\left\{\strut L'(p);\,l'(p)\leq r\right\}$.
  Hence, it is ``harder'' for a set to be a fitting set than for
  the problem without holes. Therefore, the number of fitting sets is
  smaller and is still at most $L(p)^k\leq \Omega^k$.

  As in Theorem~\ref{thm.yannakakis_poly}, the proof is an induction
  on points $p$ of $T$ (from the leaves to the root) and on fitting
  live sets $F_p\in {\cal F}_p$. 
%
  $W_{max}(p,F_{p})$ is built, for each $F_p\in {\cal F}_p$, thanks to
  dynamic programming, by ``concatenating'' some well chosen
  $W_{max}(f,F_f)$.  Given a child $f$ of $p$, we select a fitting set
  $F_f\in {\cal F}_f$ that matches $F_p$, i.e., such that $F_f\cap
  L(p)=F_p\cap L(f)$, and that maximizes the cost of $W_{max}(p,F_p)$.
  We do this for each child of $p$, and because by construction they
  match on $p$, they can be expanded to a solution
$W_{max}(p,F_p)$ that fits $T_p$. The arguments are the same as for
Theorem~\ref{thm.yannakakis_poly} and are not repeated here.
\end{proof}

We have seen that, without holes, the spill everywhere
problem on an SSA program, with few registers, is polynomial whereas
the instance with many registers ($k$) is NP-complete: the number of
spilled variables live at a given point can be arbitrarily large (up
to~$\Omega$). For a basic block, if $h$ is fixed, this is not the case
anymore. As we will see, this number is bounded by $2(h+k)$, leading
to a dynamic programming algorithm with $\cplx(|B|\Omega^{2(h+k)})$
steps.

\begin{theorem}[Dynamic programming on spilled variables]
  The spill everywhere problem with holes and many registers can
  be solved in polynomial time, for a basic block, if $h$ is fixed
  even if $w\neq 1$.
\end{theorem}

\begin{proof} 
  The key point is to first prove that, for an optimal solution, for
  each point $p$, $|L_S(p)|\leq 2(h+k)$.  Consider a point $p$ such
  that $|L_S(p)|\geq h+k+1$. We extend this point to a maximal
  interval $I$ such that on any point $p$ of this interval,
  $|L_S(p)|\geq h+k+1$. We claim that there is no spilled variable
  $v\in V_S$ completely included in~$I$. Indeed, otherwise, if $v$ were
  restored (unspilled), then, at each point $p$ of $v$, at least
  $(h+k+1)-1=h+k$ variables would have been spilled, so the register
  pressure $l'(p)\leq |L'(p)|+h\leq (\Omega-(h+k))+h=\Omega-k$ would still
  be small enough. This would contradict the optimality of the initial
  solution.  Hence, no variable of $V_S$ is completely included in~$I$:
  either it starts before the beginning of $I$, or it ends after the
  end of $I$. But $I$ is of maximal size, hence on both extremities,
  there are at most $h+k$ live spilled variables.  This means that there is
  at most $2(h+k)$ spilled variables live in any point of $I$.
  
  The rest of the proof is similar to the proofs of
  Theorems~\ref{thm.yannakakis_poly} and~\ref{thm.dynamic_holes}. The
  only difference is that spilled variables are considered instead of kept
  variables. For a point $p$, an {\em extra} live set~$E_p$ is a set
  of variables of cardinal at most $2(h+k)$ and such that, if
  $E_p$ is spilled, the new register pressure $l'(p)$ becomes lower than $r$.
  Let~${\cal E}_p$ be the set of extra sets for $p$. It has at most
  $L(p)^{2(h+k)}\leq \Omega^{2(h+k)}$ elements.

  The proof is an induction on points $p$ of $B=\{p_1, \ldots,
  p_m\}$ and on extra live sets $E_p\in {\cal E}_p$.
  Let $B_{p_i}=\{p_1, \ldots, p_i\}$. A set of variables is said to
  fit $B_p$ if, for all points in $B_p$, the register pressure
  obtained if all other variables are spilled is at most $r$. The
  induction hypothesis is that a solution $W_{max}(p,E_p)$ of maximum
  cost, that fits $B_p$, and with $L_S(p)=E_p$, can be built in
  polynomial time.
  Let $p$ be a point of $B$ and $f$ its predecessor. Let $E_{p}\in
  {\cal E}_{p}$, and an extra live set $E_{f}$ that matches
  $E_{p}$, i.e., such that $E_{f}\cap L(p)=E_p\cap L(f)$, and that
  maximizes the cost of $W_{max}(f,E_f)$. As noticed earlier,
  $\left|\strut{\cal E}_f\right|\leq \Omega^{2(h+k)}$ and it can be
  built, by induction hypothesis, in polynomial time. Because $E_p$
  and $E_f$ match, $W_{max}(f,E_{f})$ can be expanded to a solution
  $W_{max}(p,E_{p})$ that fits $B_p$. The arguments are the same as
  those used for Theorems~\ref{thm.yannakakis_poly}
  and~\ref{thm.dynamic_holes}.

  The proof is constructive and provides an algorithm based on dynamic
  programming with $\cplx(|B|\Omega^{2(h+k)})$ steps.
\end{proof}

The next two theorems show that the complexity does depend on $h$ and
$k$. If $h$ is not fixed but $k=1$, the incremental problem is
NP-complete (Theorem~\ref{thm.mincover}). If $h$ is fixed but there is
no constraints on $r$, most instances are NP-complete
(Theorems~\ref{thm.NPtwo_unweighted} and~\ref{thm.NPone_weighted}).

\begin{theorem}[From Minimum Cover]
\label{thm.mincover}
The incremental spill everywhere with holes is NP-complete even if
$w(v)=1$ for each $v\in V$ and even on a basic block, if $h$ can be arbitrary.
\end{theorem}

\begin{proof}
  The proof is a straightforward reduction from \textsf{Minimum
    Cover}~\cite[Problem SP5]{gareyjohnson}. Let $\mathcal{V}$ be
  subsets of a finite set $\mathcal{B}$ and $\mathcal{K}\leq
  |\mathcal{V}|$ be a positive integer. Does $\mathcal{V}$ contain a
  cover for~$\mathcal{B}$ of size $\mathcal{K}$ or less, i.e., a
  subset $\mathcal{V'}\subseteq\mathcal{V}$ such that every element
  of~$\mathcal{B}$ belongs to at least one member of $\mathcal{V'}$?
  Punched intervals can be seen as subsets of $B$, they contain all
  points, except chads. 

Consider an instance of Minimum Cover. To
  each element of $\mathcal{B}$ corresponds a point of $B$. To each
  element $\nu$ of $\mathcal{V}$ corresponds a punched interval $v$
  that traverses entirely $B$ and that only contains points
  corresponding to elements of $\nu$. In other words, there is a chad
  for each point not in $v$.
At each point $p$ of $B$, the number of punched intervals and chads
that contain $p$ (live variables) is exactly $\Omega=|V|$. A spilling that lowers by at
least one the register pressure $\Omega$ provides a cover of $B$ and
conversely. So, setting $K=\mathcal{K}$ and $r=\Omega-1$
proves the theorem.
\end{proof}

Notice that the previous proof is very similar to the proof of
Farach-Colton and Liberatore~\cite{Liberatore00} for Lemma~3.1. This
lemma proves the NP-completeness of the load-store optimization
problem, which is harder than our spill everywhere problem. Still,
their reduction is similar to ours since they used a trick to force
the overall load cost to be the same for all spilled variables,
independently on the number of times a variable is evicted. Hence,
the optimal solution to their load-store optimization problem just
behaves like a spill everywhere solution.

\begin{figure*}[Ht]
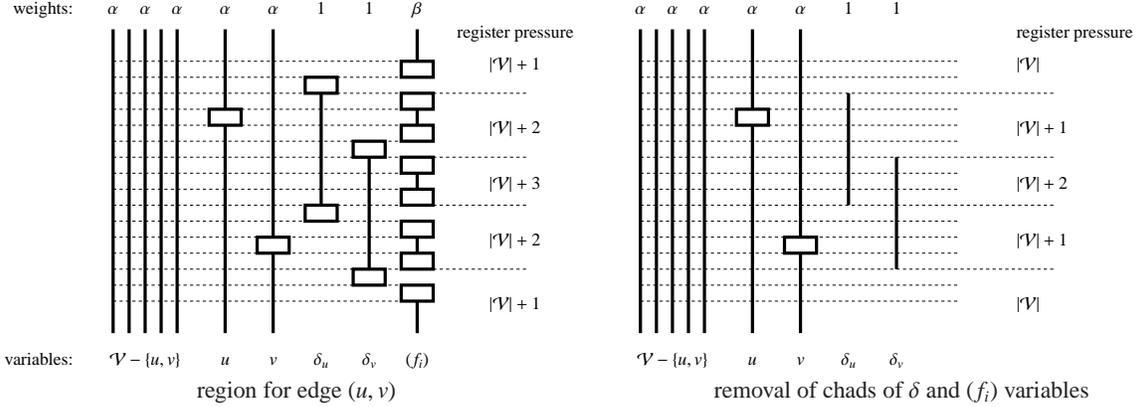

\begin{center}
\begin{tabular}{cc}
\includegraphics[height=5cm]{onechad-initial.fig} & 
\includegraphics[height=5cm]{onechad-nospill.fig}\\
region for edge $(u,v)$ & removal of chads of $\delta$ and $(f_i)$
variables
\end{tabular}
\end{center}
\caption{\label{fig.onechad-nospill}For each edge in $E$, a
  corresponding region in $B$. With $\beta$ large enough, spilling this region with $r$ registers
is equivalent to spilling the simplified region with $r-1$ registers.}
\vspace{-0.3cm}
\end{figure*}

The main limitation of the reduction used for
Theorem~\ref{thm.mincover} is that the proof needs the number of
simultaneous chads $h$ to be arbitrary large, as large as $|V|$. This
is of course not realistic for real architectures. In
practice, usually $h=2$ and even $h=1$ for paging problems.  Similarly
to ours, the reduction of Farach-Colton and Liberatore use a large
amount of simultaneous uses (in~\cite{Liberatore00} a \texttt{read}
corresponds to a use and $\alpha$ corresponds to~$h$).  Theorem~3.2
of~\cite{Liberatore00} extends their lemma to the case $\alpha=1$ but
again, it deals with load-store optimization problem, which is harder
than spill everywhere. Unfortunately, their trick cannot be
applied to prove the NP-completeness of our ``simpler'' problem and we
need to use a different reduction as shown below.

\begin{theorem}[At most $2$ simultaneous chads]
\label{thm.NPtwo_unweighted}
The spill everywhere problem with holes is NP-complete even if
$w(v)=1$ for all $v\in V$, even with at most $2$
simultaneous chads, and even on a basic block.
\end{theorem}

\begin{proof}
  The proof is a straightforward reduction from \textsf{Independent
    Set}~\cite[Problem GT20]{gareyjohnson}. Let $G=(\mathcal{V},E)$ be
  a graph and $\mathcal{K}\leq |\mathcal{V}|$ be a positive integer.
  Does $G$ contain an independent set (stable) $\mathcal{V}_S$ of size
  $\mathcal{K}$ or more, i.e., a subset $\mathcal{V}_S\subseteq \mathcal{V}$
  such that $|\mathcal{V}_S|\geq \mathcal{K}$ and no two
  vertices in $\mathcal{V}_S$ are joined by an edge (adjacent) in $E$?

  Consider an instance of Independent Set. To each vertex $\nu
  \in\mathcal{V}$ of $G$ corresponds a variable $v\in V$ which is live
  from the entry of~$B$ to its exit. To each edge $(\mu,\nu)\in E$ of
  $G$ corresponds a point~$p(u,v)$ of $B$ that contains a use of the
  corresponding variables~$u$ and $v$. In other words, there are two
  chads for each point of $B$.  The key point is to notice that
  spilling $K$ variables in $V_S$ lowers the register pressure to
  $|V|-K+1$ if and only if the corresponding set of
  vertices~$\mathcal{V}_S$ is an independent set. Indeed, if
  $\mathcal{V}_S$ contains two adjacent vertices $u$ and $v$, then at
  point $p(u,v)$, the register pressure would be $|V|-K+2$.  Hence,
  by letting $K=\mathcal{K}$ and $r=|V|-K+1$, we get the desired
  reduction. Indeed, if there exist $k\leq K$ variables that, when
  spilled, lead to a register pressure at most $r=|V|-K+1$ then, first,
  $k$ must be equal to $K$ and, second, the corresponding vertices form
  an independent set of size $K$. Conversely, if there is an
  independent set of size at least $K$, then spilling the
  corresponding variables leads to a register pressure at most
  $|V|-K+1$.
\end{proof}

\begin{theorem}[No simultaneous chads]
\label{thm.NPone_weighted}
The spill everywhere problem with holes is NP-complete even if $h=1$
and for a basic block.
\end{theorem}

\begin{figure*}[Ht]
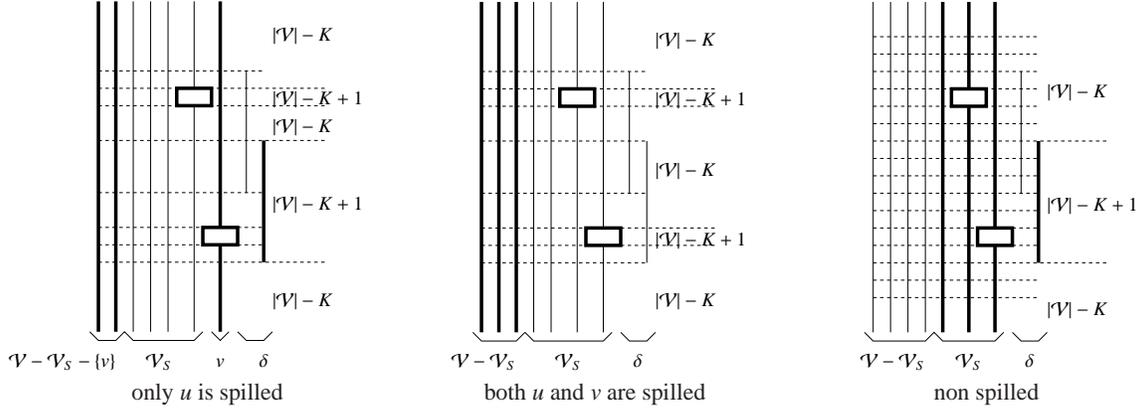

\begin{center}
\begin{tabular}{p{0.5cm}cp{1.1cm}cp{1.1cm}c}
&\includegraphics[height=5cm]{onechad-u.fig} & &
\includegraphics[height=5cm]{onechad-uv.fig}& &
\includegraphics[height=5cm]{onechad.fig}\\
&only $u$ is spilled && both $u$ and $v$ are spilled && non spilled
\end{tabular}
\end{center}
\caption{\label{fig.onechad}Different configurations whether $u$ and
  $v$ are spilled or not with $r=|\mathcal{V}|-K+1$ registers. Non
  spilled variables are in bold.}
\vspace{-0.3cm}
\end{figure*}

\begin{proof}
  As for Theorem~\ref{thm.NPtwo_unweighted}, the proof is a reduction
  from \textsf{Independent Set}. Consider an instance of Independent
  Set. To each vertex $\nu \in\mathcal{V}$ of $G$ corresponds a
  variable $v\in V$ (called vertex variables), which is live from the
  entry of $B$ to its exit.  To each edge $(\mu,\nu)\in E$ of $G$
  corresponds a region in $B$ where $u$ and $v$ are consecutively
  used. As depicted in Figure~\ref{fig.onechad-nospill}, such a region
  contains two additional overlapping local variables $\delta_u$ and
  $\delta_v$ (called $\delta$ variables). For real codes, every live
  range must contain a chad at the beginning and a chad at the end.
  For our proof, we need to be able to remove the complete live range
  of a $\delta$ variable, which is not possible because of the
  presence of chads for such variables. To avoid this problem, we
  increase the register pressure by $1$ everywhere, except
  where~$\delta$ variables have chads. See
  Figure~\ref{fig.onechad-nospill} again: we add new variables $f_i$
  such that the union of their live ranges covers exactly all points
  of~$B$, except the points that correspond to the chad of a $\delta$
  variable. The cost $\beta$ of spilling a variable~$f_i$ will be
  chosen large enough so that $f_i$ variables are never spilled in an
  optimal solution.  So, from now on, without loss of generality, we
  consider the simplified version of the region (right hand side of
  Figure~\ref{fig.onechad-nospill}) where $\delta$ live ranges contain
  no chads. We let $K=\mathcal{K}$ and $r=|\mathcal{V}|-K+1$. The cost
  for spilling a vertex variable is $\alpha$ while the cost for
  spilling a $\delta$ variable is 1. The suitable value for $\alpha$ will be
  determined later.

  The trick is to make sure that an optimal solution of our spilling
  problem spills exactly $K$ vertex variables and at least $|E|$
  of the $\delta$ variables (one per region). We do so by letting
  $\alpha=2|E|+1$ (in fact $\alpha=|E|+1$ would be enough but we do so
  to simplify the proof). First, spilling $K-1$ vertex variables in
  addition to all $\delta$ variables is not enough: on the chad of
  one of the spilled variables, the register pressure will be lowered
  to $|\mathcal{V}|-(K-1)+1=|\mathcal{V}|-K+2>r$. Second, spilling
  $K$ vertex variables requires to spill at least one $\delta$
  variable per region and spilling all $\delta$ variables is enough.
  Hence, the minimum cost of a spilling with exactly $K$ vertex
  variables is between $K\alpha+E$ and $K\alpha+2E$. Finally, spilling
  $K+1$ vertex variables has a cost equal to $(K+1)\alpha=K\alpha+2|E|+1$.

  Now, it remains to show that the cost of an optimal spilling is
  $K\alpha+E$ if and only if the spilled variables define an
  independent set for $G$. Consider an edge $(u,v)$. All situations
  are depicted in Figure~\ref{fig.onechad}. If both $u$ and $v$ are
  spilled (in this case, $\mathcal{V}$ is not a
  stable set), then both $\delta_u$ and $\delta_v$ must be spilled
  and the cost cannot be $K\alpha+E$. Otherwise, spilling either
  $\delta_u$ or $\delta_v$ is enough.
\end{proof}

\section{Conclusion}
\label{sec_conclusion}

Recent results on the SSA form have opened promising directions for the design
of register allocation heuristics, especially for dynamic embedded
compilation. Studying the complexity of the spill everywhere problem
was important in this context. Unfortunately, our work shows that
SSA does not simplify the spill problem like it does for the
assignment (coloring) problem. Still, our
results can provide insights for the design of
aggressive register allocators that trade compile time for
provably ``optimal'' results.
Our study considers different singular variants of the
spill everywhere problem.
\begin{enumerate}
\item We distinguish the problem without or with holes depending on
  whether use operands of instructions can reside in memory slots or
  not. Live ranges are then contiguous or with chads. 
\item For the variant with chads, we study the influence of
  the number of simultaneous chads (maximum number of use operands of
  an instruction and maximum number of definition operands of an
  instruction). 
\item We distinguish the case of a basic block (linear sequence)
and of a general SSA program (tree). 
\item Our model uses a cost function for spilling a variable. We
  distinguish whether this cost function is uniform (unweighted) or
  arbitrary (weighted).
\item Finally, in addition to the general case, we consider the singular case of 
  spilling with few registers and the case of an incremental spilling that 
  would lower the register pressure one by one.
\end{enumerate}
The classical furthest-first greedy algorithm is optimal only for the
unweighted version without holes on a basic block. An ILP formulation
can solve, in polynomial-time, the weighted version, but
unfortunately, only for a basic block, not a general
SSA program.


The positive result of our study for architectures with few registers
is that the spill everywhere problem with a bounded number of
registers is polynomial even with holes. Of course, the complexity is
exponential in the number of registers, but for architectures like
x86, it shows that algorithms based on dynamic programming can be
considered in an aggressive compilation context. 
In particular, it is a possible alternative to commercial solvers required by ILP
formulations of the same problem.
%
For architectures with a large number of registers, we have studied the
{\it a priori} symmetric problem where one needs to decrease the
register pressure by a constant number. Our hope was to design a
heuristic that would incrementally lower one by one the register
pressure to meet the number of registers.  Unfortunately,  this
problem is NP-complete too.

To conclude, our study shows that complexity also comes from the
presence of chads. The problem of spill everywhere with chads is
NP-complete even on a basic block.  On the other hand, the incremental
spilling problem is still polynomial on a basic block provided that
the number of simultaneous chads is bounded. Fortunately, this number
is very low on most architectures.



\acks
We would like to thank Christophe Guillon and Sebastian Hack for 
fruitful discussions.



\end{document}